\documentclass[pra,aps,twocolumn]{revtex4-1}

\usepackage[utf8x]{inputenc}
\usepackage{color}
\usepackage{bbm} 

\usepackage{amsfonts,amsmath,amssymb,stmaryrd}

\usepackage{graphicx}
\usepackage{subfigure}  
\usepackage{bbm}
\usepackage{hyperref}
\usepackage{epsfig}
\usepackage{mathrsfs}
\usepackage{verbatim}
\usepackage{ulem}	
\usepackage{siunitx}
\usepackage{soul} 

\usepackage[all]{hypcap}
\usepackage{braket}

\newcommand{\hlf}{\mbox{$\frac{1}{2}$}}
\newcommand{\qtr}{\mbox{$\frac{1}{4}$}}
\newcommand{\tth}{\mbox{$\frac{2}{3}$}}

\newcommand{\sig}{\hat{\sigma}}

\newcommand{\av}[1]{\langle #1 \rangle}
\renewcommand{\H}{\mathcal{H}}
\newcommand{\hrho}{\hat{\rho}}
\newcommand{\SA}{\mathrm{SA}}

\newcommand{\dt}{\partial_t}

\usepackage{array}

\begin{document}
\normalem	

\title{Many-body dynamics of holes in a driven, dissipative spin chain \\
of Rydberg superatoms}

\author{Fabian Letscher}
\affiliation{Department of Physics and Research Center OPTIMAS, 
University of Kaiserslautern, D-67663 Kaiserslautern, Germany}
\affiliation{Graduate School Materials Science in Mainz, 
Gottlieb-Daimler-Strasse 47, D-67663 Kaiserslautern, Germany}

\author{David Petrosyan}
\affiliation{Institute of Electronic Structure and Laser, FORTH, 
GR-71110 Heraklion, Crete, Greece}

\author{Michael Fleischhauer}
\affiliation{Department of Physics and Research Center OPTIMAS,
D-67663 Kaiserslautern, University of Kaiserslautern, Germany}

\begin{abstract}
Strong dipole-dipole interactions between atoms in high-lying Rydberg 
states can suppress multiple Rydberg excitations within a micron-sized 
trapping volume and yield sizable Rydberg level shifts at larger 
distances. Ensembles of atoms in optical microtraps then form 
Rydberg superatoms with collectively enhanced transition rates 
to the singly excited state. 
These superatoms can represent mesoscopic, strongly-interacting spins.
We study a regular array of such effective spins driven by a laser field 
tuned to compensate the interaction-induced level shifts between neighboring 
superatoms. During the initial transient, a few excited superatoms seed
a cascade of resonantly facilitated excitation of large clusters
of superatoms. Due to spontaneous decay, the system then relaxes to 
the steady state having nearly universal Rydberg excitation density  
$\rho_{\mathrm{R}} = 2/3$.
This state is characterized by highly-nontrivial equilibrium dynamics 
of quasi-particles -- excitation holes in the lattice of Rydberg excited 
superatoms. 
We derive an effective many-body model that accounts for hole mobility 
as well as continuous creation and annihilation of holes upon collisions 
with each other. We find that holes exhibit a nearly incompressible 
liquid phase with highly sub-Poissonian number statistics and finite-range 
density-density correlations.
\end{abstract}

\date{\today}

\maketitle

\section{Introduction}
Strongly-interacting many body systems subject to external driving
and coupled to (possibly tailored) reservoirs offer a new route to create 
and stabilize interesting states of matter. As a simple example, a quantum 
state can be made immune to particle losses if it is the stationary state 
of an open system coupled to a particle reservoir. 
Furthermore, the competition between coherent driving and dissipation 
can lead to exotic steady states \cite{Kapit2014, Vermersch2016} 
and phase transitions in open many body systems  
\citep{Kessler2012, Lang2015, Mendoza-Arenas2015, Wilson2016, Marcuzzi2016d, Casteels2017a}. 

Rydberg atoms are well suited to study the interplay between strong 
interaction and coupling to coherent laser fields and dissipative 
environments. They are thus prime candidates to investigate 
many-body physics of driven, dissipative spin models.
A prominent and well studied consequence of the strong, long-range interaction 
between atoms in Rydberg states is the so called blockade phenomenon, 
whereby a Rydberg excited atom suppresses further excitations 
within a certain blockade distance \cite{Lukin2001}. Rydberg blockade 
in a dilute gas or in a regular array of single atoms leads to short-range 
spatial ordering of excitations, as was studied theoretically 
\cite{Ates2012a,Hoening2013,Petrosyan2013,Petrosyan2013c} and demonstrated 
experimentally \cite{Schauss2012, Labuhn2015}. In the so-called anti-blockade 
regime, successive excitation of atoms at a certain distance from each other 
can be resonantly enhanced 
\cite{Ates2012, Lesanovsky2014, Schempp2014, Urvoy2015, Marcuzzi2017, Carr2013, Malossi2014},
which lead to the lively debate on the possibility 
of attaining bistable steady states \cite{Carr2013, Malossi2014, Weimer2015, Sibalic2016, Letscher2016}.

When many atoms are confined within the blockade distance from 
each other, they form an effective two-level system -- Rydberg 
superatom -- that can accommodate at most one Rydberg excitation 
\cite{Lukin2001, Honer2011a, Petrosyan2013}. The coupling of a
superatom to the laser radiation is collectively enhanced,  
while the steady state probability of a single Rydberg excitation can 
approach unity. This permits the level of control of single collective 
spins represented by superatoms far exceeding that for individual atoms. 
Moreover, being composed of many atoms, superatoms are relatively 
insensitive to atom number fluctuations and losses.
Regular arrays of spins represented by superatoms can then be prepared 
with less experimental effort, which should be contrasted with the 
sophisticated dynamical preparation techniques used to realize 
defect-free arrays of individual Rydberg atoms \cite{Barredo2016,Endres2016}. 
This, together with the strong, long-range
interactions between the Rydberg excitations, makes superatoms ideal 
building blocks for realizing dissipative many-body spin models and 
analyzing their dynamics.

\begin{figure}[t]
  \centerline{\includegraphics[width=0.85\columnwidth]{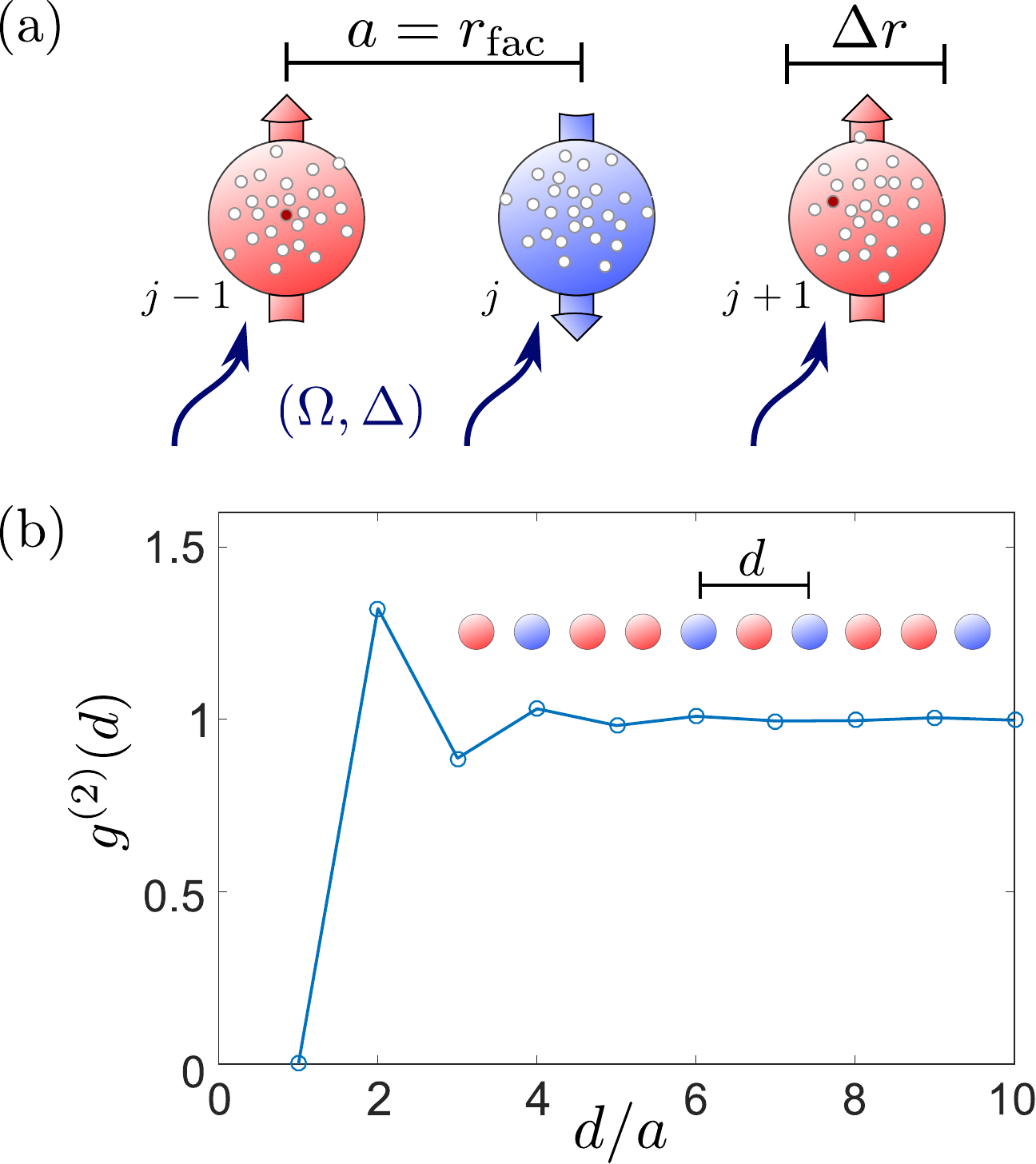}}
  \caption{(a) Schematics of the chain of effective spins represented by 
    Rydberg superatoms separated from each other by the lattice constant $a$. 
    Each superatom $j$ contains $N$ atoms confined within a microtrap 
    of linear dimension $\Delta r \ll a, a_\mathrm B$, with $a_\mathrm B$ 
    being the Rydberg blockade distance.  
    Atoms in the ground state (open dots) are excited to 
    the Rydberg state (red filled dots) by a uniform laser field 
    with Rabi frequency $\Omega$ and detuning $\Delta$. 
    We tune $\Delta$ to compensate the interaction-induced level shift
    of Rydberg states of neighboring superatoms leading to resonantly
    facilitated excitation at distance $r_\textrm{fac}=a$. 
    (b) In the steady state of a continuously driven lattice of superatoms, 
    having nearly universal density $\rho_{\mathrm{R}}=2/3$ of Rydberg 
    excitations (red filled circles), 
    the typical two-particle correlation function $g^{(2)}(d)$ for 
    the excitation holes (blue filled circles) corresponds to a 
    liquid of hard rods of length $2a$.} 
  \label{fig:SuperatomLattice}
\end{figure}

Single Rydberg superatoms have been observed in several experiments 
\cite{Heidemann2007, Dudin2012a, Ebert2015, Weber2015, Zeiher2015, Labuhn2015}. 
A two-dimensional square lattice of superatoms with nearest-neighbor 
excitation blockade was studied in \cite{Hoening2014}, demonstrating 
the possibility of a phase transition to an anti-ferromagnetic steady state 
with spontaneously broken lattice symmetry. In the complementary 
interaction regime of the Rydberg anti-blockade, little is known about 
the many-body dynamics and the steady state of a lattice of superatoms. 
Here we study a one-dimensional lattice of Rydberg superatoms 
[see Fig.~\ref{fig:SuperatomLattice}(a)], in which an already excited 
superatom facilitates resonant excitation of its neighbor, but the presence 
of two excited neighbors suppresses the excitation. This systems exhibits 
interesting excitation dynamics and a highly-nontrivial steady state 
characterized by an almost universal density $\rho_{\mathrm{R}}=2/3$ 
of Rydberg excitations with strongly suppressed number fluctuations. 
We show that this behavior can be explained in terms of mobile 
excitation holes on the background of Rydberg excited lattice. 
The holes behave as a nearly incompressible liquid of hard rods with 
characteristic two-particle correlations 
[see Fig.~\ref{fig:SuperatomLattice}(b)]. 
We derive and verify an effective many-body model for holes. 
Varying the parameters of the effective model, we find a cross-over 
between a liquid of holes with density-density correlations 
peaked at the distance of two lattice periods, $2a$, and the onset of 
crystalline order with period $3a$. In both cases the density of holes 
is $\rho_h=1/3$ with highly suppressed number fluctuations. 

The paper is organized as follows. 
In Sec. \ref{sec:SuperatomLattice}, we formulate the model for a regular 
array of superatoms and derive the formalism for the efficient treatment 
of the system. 
In Sec. \ref{sec:Results} we present the results of numerical 
simulations of the dynamics of the chain of superatoms (driven spin chain) 
and introduce the effective hole model that leads to an intuitive physical 
picture for the equilibrium phase of the system.
Section~\ref{sec:Conclusion} summarizes our results. 
Technical derivations are deferred to the Appendices \ref{ap:SAREM},
\ref{ap:AntiBlockadeConditions}, and \ref{ap:HoleDynamicsModel}.


\section{Chain of Rydberg superatoms}
\label{sec:SuperatomLattice}

In this Section, and in Appendices \ref{ap:SAREM} and 
\ref{ap:AntiBlockadeConditions}, starting from the fully 
quantum many-body master equation, we derive rate equations for 
the chain of laser-driven and mutually interacting Rydberg superatoms.
These rate equations will then be used in Sec.~\ref{sec:Results}
for numerical simulations of the dynamics and steady state of the
many-body system.

\subsection{The microscopic model}

We consider an ensemble of cold atoms trapped in a regular array of 
microtraps \cite{Dumke2002, Wuertz2009, Weber2015, Ebert2015, Nogrette2014}
or a long-wavelength optical lattice with the period $a$ of a few microns. 
Each lattice site $j$ contains on average $N$ atoms, 
see Fig. \ref{fig:SuperatomLattice}(a). 
A laser field of carrier frequency $\omega$ drives the atoms on the transition 
from the ground state $\ket{g}$ to the excited Rydberg state $\ket{e}$ with 
the Rabi frequency $\Omega$ and detuning $\Delta= \omega - \omega_{eg}$.
In the frame rotating with frequency $\omega$, the coherent excitation 
dynamics of the atoms is described by the Hamiltonian
($\hbar = 1$)
\begin{equation}
\label{eq:Hamiltonian}
\H = \sum_{k} \left[ \Omega \left( \sig_{eg}^{k} 
+ \sig_{ge}^{k} \right) - \Delta \sig_{ee}^{k}  \right] 
+ \sum_{k \neq k'} V(\vec{r}_{k},\vec{r}_{k'}) \,
\sig_{ee}^{k} \otimes \sig_{ee}^{k'} , \quad 
\end{equation}
where $\sig_{\mu \nu}^{k} \equiv \ket{\mu}_k \! \bra{\nu}$ is the transition
($\mu \neq \nu$) or projection ($\mu = \nu$) operator for the $k$th atom, 
and $V(\vec{r}_{k},\vec{r}_{k'}) = \frac{C_p}{|\vec{r}_{k}-\vec{r}_{k'}|^p}$
is the interaction potential between pairs of atoms at positions 
$\vec{r}_{k},\vec{r}_{k'}$ excited to the Rydberg state $\ket{e}$. 
The usual van der Waals interaction corresponds to $p=6$. 

Atoms excited to the Rydberg state $\ket{e}$ spontaneously decay 
to the ground state with the rate $\Gamma_s$, and are dephased with 
a typically much larger rate $\Gamma_d$ due to collisions of the 
Rydberg electron with the ground state atoms \cite{Balewski2013, Gaj2014}, 
atomic motion in the inhomogeneous trapping potential,  
or intermediate state decay if $\ket{g} \to \ket{e}$ is a two-photon 
transition \cite{Schauss2012, Schauss2015}. 
The dissipative dynamics is described 
by the master equation for the density matrix $\hrho$ of the system,
\begin{equation}
\label{eq:MasterEquation}
\dt \hrho = -i[\H,\hrho] 
+ \sum_k \sum_{\mu =s,d} 
\left( L_\mu^k \hrho L^{k \dagger}_\mu 
- \hlf \{ L^{k \dagger}_\mu L_\mu^{k}, \hrho \} \right)  ,
\end{equation}
where $L_\mu^k$ are the Lindblad jump operators for the two relaxation 
processes assumed acting independently on each atom $k$ as 
$L_s^{k} = \sqrt{\Gamma_s}\sig_{ge}^{k}$ and
$L_d^{k} = \sqrt{\Gamma_d} \sig_{ee}^{k}$. 

\subsection{Rate equations for Rydberg superatoms}

\begin{figure}[t]
  \centerline{\includegraphics[width=1.0\columnwidth]{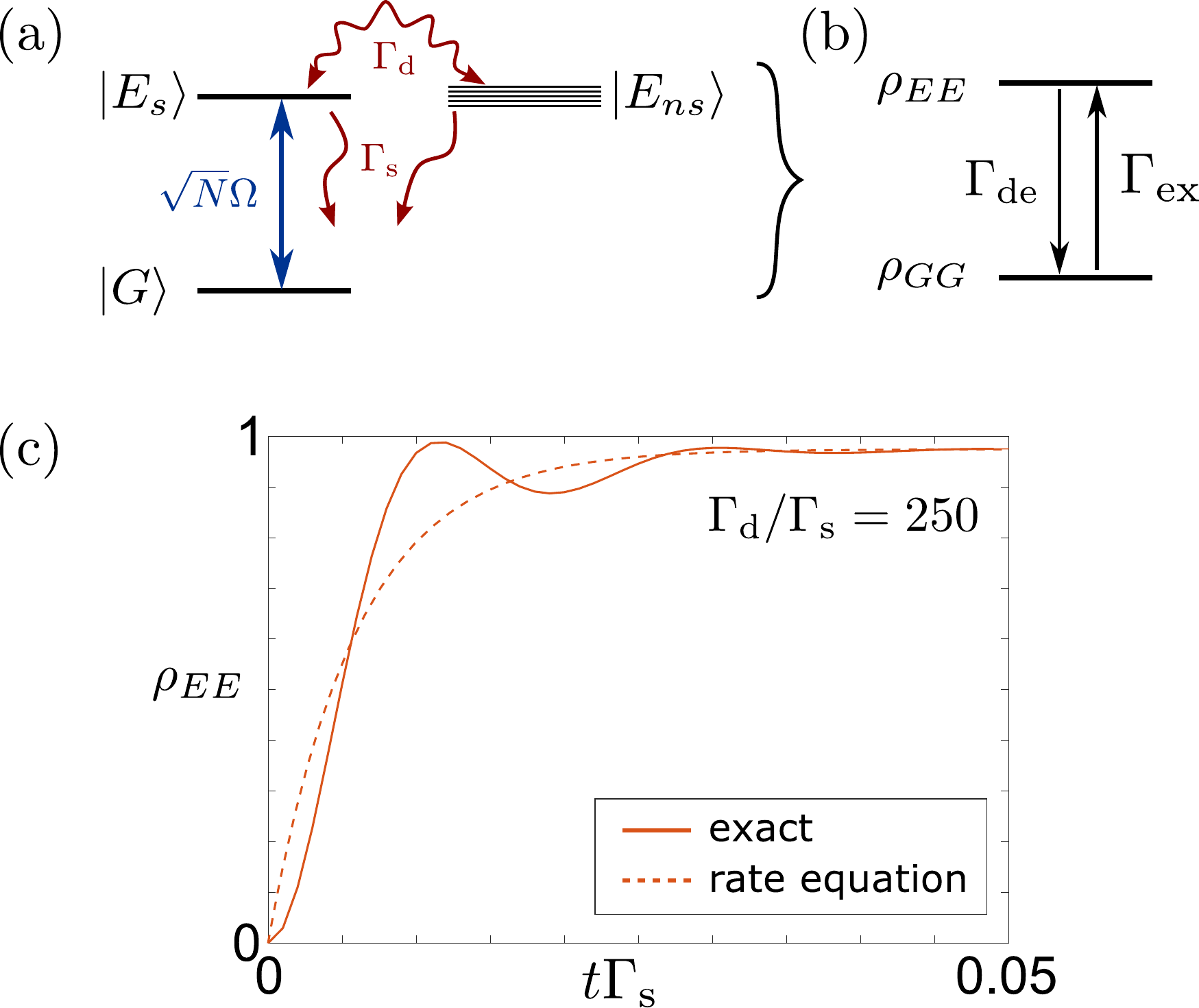}}
  \caption{(a)~Level scheme of a single Rydberg superatom consisting 
    of $N$ atoms. The laser field couples the collective 
    ground state $\ket{G}$ to the symmetric single excitation 
    state $\ket{E_s}$ with the Rabi frequency $\sqrt{N} \Omega$.
    Dephasing of the Rydberg state with rate $\Gamma_d$ leads 
    to population of $(N-1)$ nonsymmetric single
    excitation states $\ket{E_{ns}}$. The single excitation states
    decay spontaneously to the ground state with rate $\Gamma_s$.
    States with multiple excitations are not populated 
    due to the strong Rydberg blockade. 
    (b)~In the approximate rate equation model,
    the Rydberg superatom is excited and de-excited 
    with the corresponding rates $\Gamma_\mathrm{ex}$ 
    and $\Gamma_\mathrm{de}$ which depend on the effective 
    detuning $\Delta_{\mathrm{eff}}$. 
    (c) Comparison of the excitation dynamics of a single superatom,
    containing $N = 50$ atoms, as obtained with the rate equation 
    model (dashed line) and from the exact solution of the master 
    equation for the density operator (solid line). 
    The single-atom Rabi frequency $\Omega/\Gamma_s = 25$ 
    and dephasing rate $\Gamma_d/\Gamma_s = 250$.  } 
  \label{fig:SuperatomLevelScheme}
\end{figure}

In Eq.~(\ref{eq:Hamiltonian}) we can split the sum over all the atoms 
into two parts: the sum over the lattice sites $j$, and the sum over 
the atoms $k_j$ in each lattice site. We assume that all $N$ atoms
within each lattice site are confined within a small spatial interval 
$\Delta r \ll a$, such that the interatomic interaction energy 
$C_p/(\Delta r)^p$ exceeds all the relevant energy scales pertaining to 
the atoms, namely, the laser Rabi frequency $\Omega$ and detuning $\Delta$, 
as well as the atomic spontaneous decay $\Gamma_s$ and dephasing $\Gamma_d$
rates and the resulting Rydberg-state excitation linewidth is
$w \simeq 2 \Omega \sqrt{\gamma/\Gamma_s}$ with 
$\gamma \equiv \frac{1}{2}(\Gamma_d + \Gamma_s)$ \cite{Petrosyan2013}. 
This allows us to neglect all the multi-atom states containing more than 
one Rydberg excitation per lattice site \cite{Lukin2001, Petrosyan2013}. 
If on the spatial scale $\Delta r \lesssim 1\:\mu$m of such a Rydberg 
superatom the laser field can be assumed uniform, it would couple the 
collective ground state $\ket{G} \equiv \ket{g_{1} g_2 \ldots g_{N}}$ only to 
the symmetric single excitation state $\ket{E_s} \equiv \frac{1}{\sqrt{N}} 
\sum_k \ket{g_1 g_2 \ldots e_k \ldots g_{N}}$ with the collectively 
enhanced Rabi frequency $\sqrt{N} \Omega$,
see Fig.~\ref{fig:SuperatomLevelScheme}(a). There are, in addition,
$(N-1)$ nonsymmetric single excitation states $\ket{E_{ns}}_m$, 
labeled by index $m$, decoupled from the laser field. 
We can then recast the Hamiltonian as
\begin{multline}
\H \simeq \sum_j \left[ \sqrt{N} \Omega 
\Bigl(\ket{G^j}\bra{E_s^j} + \ket{E_s^j}\bra{G^j}\Bigr) \right]
- \Delta \hat{P}_{EE}^j \\
+ \sum_{j>i} \frac{C_p}{|\vec{r}_j-\vec{r}_i|^p} 
\hat{P}_{EE}^j \otimes \hat{P}_{EE}^i,
\end{multline}
where $\hat{P}_{EE}^j \equiv \ket{E_s^j}\bra{E_s^j} 
+ \sum_m \ket{E_{ns}^j}_{m} \! \bra{E_{ns}^j}$ is the projector onto the manifold 
of $N$ single Rydberg excitation states of superatom at site $j$, 
and $\vec{r}_j = \frac{1}{N} \sum_{k_j} \vec{r}_{k_j}$ is its center of mass 
coordinate. 

Within each Rydberg superatom, the dephasing couples 
incoherently the symmetric single excitation state $\ket{E_s}$ 
to all $(N -1)$ nonsymmetric states $\ket{E_{ns}}$ with the rate 
$\Gamma_d (N -1)/N$ which approaches $\Gamma_d$ for $N \gg 1$. 
In turn, the reverse coupling of the nonsymmetric states 
to the symmetric state has the rate $\Gamma_d /N$. All 
single excitation states decay back to the ground state $\ket{G}$ 
with rate $\Gamma_s$, see Fig.~\ref{fig:SuperatomLevelScheme}(a). 
In Appendix \ref{ap:SAREM} we derive a simple rate equation model 
describing the excitation dynamics of a superatom in the limit 
of strong dephasing $\Gamma_d \gtrsim \Omega$. Starting from 
the density matrix equations for a single Rydberg superatom, 
we adiabatically eliminate all coherences and the population of 
the symmetric excited state, which scales as $\sim 1/N$. 
The superatom then reduces to an effective two level system, 
see Fig.~\ref{fig:SuperatomLevelScheme}(b), and its dynamics 
is governed by the rate equations for the populations of the 
ground $\rho_{GG}^j$ and excited $\rho_{EE}^j$ states,  
\begin{eqnarray}
 \frac{\partial}{\partial t}\rho_{EE}^j 
&=& \Gamma_\mathrm{ex}(\Delta_j) \rho_{GG}^j 
- \Gamma_\mathrm{de}(\Delta_j) \rho_{EE}^j, \label{eq:rate-eq}\\
&& \rho_{GG}^j + \rho_{EE}^j \simeq 1,\nonumber
\end{eqnarray}
where the excitation and de-excitation rates are given by 
\begin{subequations}
\begin{align}
\label{eq:Gammaexde}
\Gamma_\mathrm{ex}(\Delta_j) &= 
\frac{(N-1)\, \chi(\Delta_j)}
{N \, \chi(\Delta_j)+ 2 \gamma} \Gamma_d, \\
\Gamma_\mathrm{de}(\Delta_j) &= 
\frac{\chi(\Delta_j)}
{N\, \chi(\Delta_j) + 2 \gamma} \Gamma_d + \Gamma_s,
\end{align}
\end{subequations}
with $\chi(\Delta_j) \equiv \frac{2 \Omega^2 \gamma}{\gamma^2 + (\Delta_j)^2}$.
Here
\begin{equation}
\Delta_j = \Delta - \sum_{i\neq j} 
\frac{C_p}{|\vec{r}_i-\vec{r}_j|^p} \rho_{EE}^i 
\label{eq:Deltaeff}
\end{equation}
is the effective detuning of superatom $j$ which includes the Rydberg level 
shift due to the interaction with all the other superatoms in the Rydberg state.
In Fig.~\ref{fig:SuperatomLevelScheme}(c), and in more detail in 
Appendix~\ref{ap:SAREM}, 
we compare the dynamics of a single superatom as obtained 
from the exact solution of the complete set of the density matrix equations 
and the solution of the rate equations. We observe that the rate equation model
approximates well the relaxation timescale of a superatom and the steady 
state population of the excited state, 
$\rho_{EE} = \frac{\Gamma_\mathrm{ex}}{\Gamma_\mathrm{ex}+\Gamma_\mathrm{de}}$.
Remarkably, unlike for a single two-level atom, the excitation probability 
of the superatom under continuous (near-)resonant driving and in the presence 
of strong dephasing $\gamma \gtrsim \Omega$ can approach unity, 
$\rho_{EE} \simeq  \frac{N}{N+1} \to 1$, with increasing $N$.  

\subsection{Facilitated excitation of superatoms}
\label{sec:AntiBlockade}

The laser irradiates continuously and uniformly the 1D chain of superatoms.
We set the detuning $\Delta$ of the laser to be equal to the interaction 
strength $V(a) = C_p/a^p \equiv V_{\mathrm{N}}$ between neighboring superatoms, 
$\Delta = V_{\mathrm{N}}$. We neglect the interaction between the next to 
nearest neighbors. The conditions for the validity of our treatment of the 
chain of superatoms are discussed in Appendix~\ref{ap:AntiBlockadeConditions}. 
In this so-called facilitation regime \cite{Lesanovsky2014}, an already excited 
superatom shifts the frequency of its nearest neighbor into resonance with the laser. 
The excitation $\Gamma_\uparrow$ and de-excitation $\Gamma_\downarrow$ rates for 
the facilitated superatom, having one and only one, excited neighbor, are 
\begin{subequations}
\begin{align}
\Gamma_\uparrow &= \Gamma_\mathrm{ex}(\Delta_\mathrm{eff} = 0),\\
\Gamma_\downarrow &= \Gamma_\mathrm{de}(\Delta_\mathrm{eff} = 0).
\end{align}
\end{subequations}
On the other hand, a superatom surrounded by non-excited neighbors 
is non-resonant with the laser and has a much lower excitation rate
\begin{equation}
\Gamma_\mathrm{seed} = \Gamma_\mathrm{ex}(\Delta_\mathrm{eff} = \Delta) .
\end{equation}
But once excited, it will play the role of a seed for a rapid growth 
of a cluster of excited superatoms. Naively, such clusters will grow 
until they collide and nearly all of the superatoms in a lattice will
be excited. However, spontaneous decay of superatoms with rate $\Gamma_s$ 
will produce excitation holes -- ground state superatoms surrounded 
by excited superatoms. A hole cannot be resonantly excited as its 
Rydberg state is shifted by the interaction with two excited neighbors 
by $2 V_{\mathrm{N}}$, leading to the effective detuning 
$\Delta_\mathrm{eff} = \Delta -2 V_{\mathrm{N}} = - \Delta$. Hence, 
the rate to refill the hole turns out to be the same 
highly-suppressed seed rate $\Gamma_\mathrm{seed}$.   

In the facilitation regime that we consider, the typical hierarchy of 
the relevant rates is 
$\Gamma_\mathrm{seed} \ll \Gamma_s,\Gamma_\mathrm{\downarrow} 
\ll \Gamma_\mathrm{\uparrow}$.

\section{The many-body dynamics}
\label{sec:Results}

Upon turning on the excitation laser, the chain of superatoms under the 
facilitation conditions described above will exhibit transient dynamics
on the timescale of $t \sim \Gamma_s^{-1}$ and then settle in a steady state. 
The excitation transients will be described later in this Section.
First we discuss the steady-state characterized by the average
Rydberg excitation density of $\rho_\mathrm{R} \simeq 2/3$ and highly 
nontrivial equilibrium dynamics of the excitation holes. 

\subsection{Steady-state distribution of holes}
\label{sec:HoleDynamicsSS}

\begin{figure}
\label{fig:HoleDynamics}
\centerline{\includegraphics[width=1.0\columnwidth]{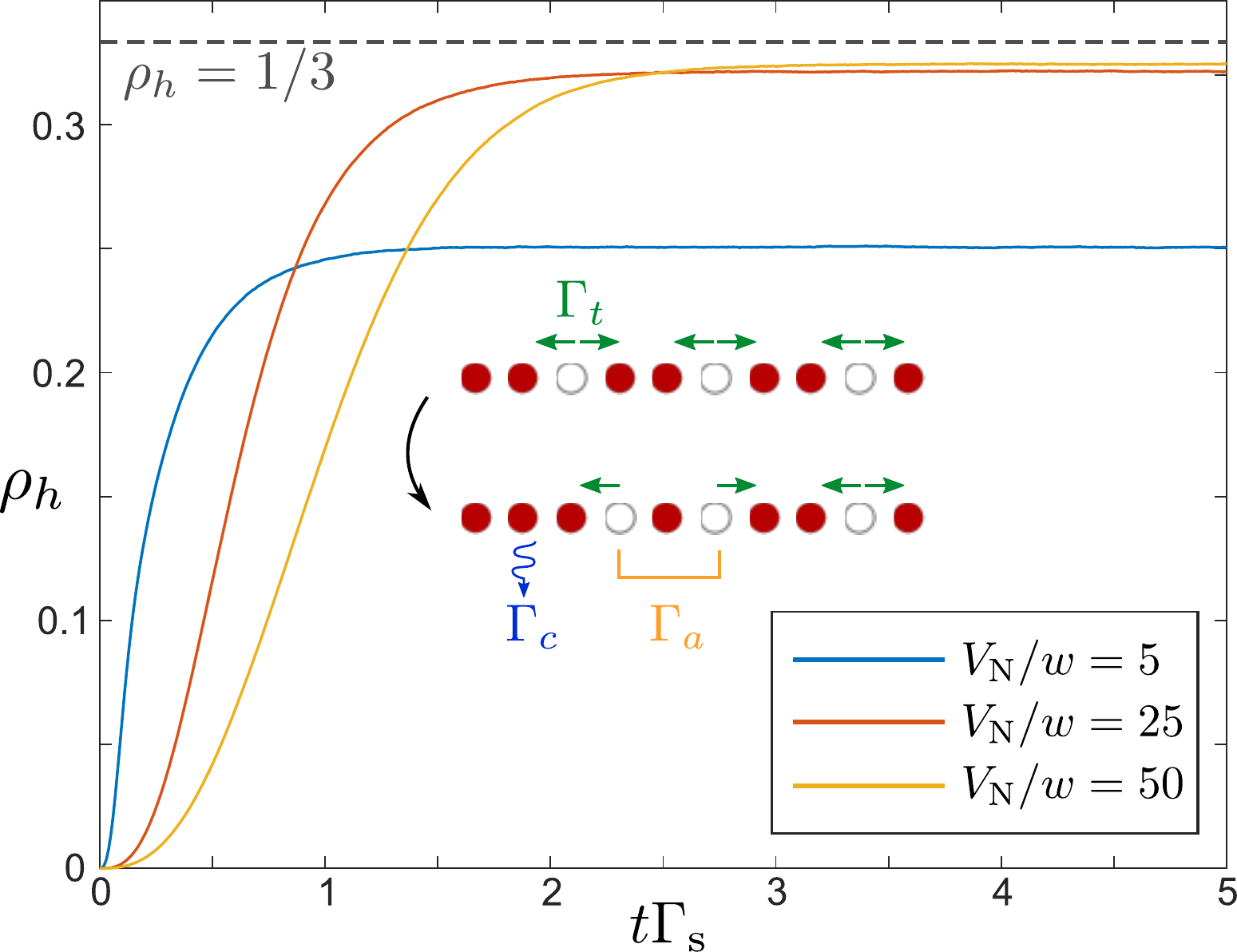}}
\caption{Density of excitation holes $\rho_h$ versus excitation 
time obtained from the rate equation simulations for various interaction 
strength $V_\mathrm{N}$. Parameters are $N=50$, $\Omega = 25 \Gamma_s$, 
and $\Gamma_d=250\Gamma_s$, leading to a single-atom excitation
linewidth $w \simeq 570 \Gamma_s$. 
The simulations are performed for a chain of $L=5000$ superatoms  
with one initial seed excitation and averaged over $500$ realizations 
of the dynamics. 
Inset: Schematics of the effective model for hole dynamics, as described
in Sec.~\ref{sec:HoleModel}. 
Holes are created with rate $\Gamma_c$, annihilated with rate $\Gamma_a$ 
and transported (hop from site to site) with rate $\Gamma_t$.}
\end{figure}

By definition, a hole is a ground state superatom surrounded by two 
excited superatoms. Holes originate from collisions of growing clusters 
of Rydberg excitations during the transient and spontaneous 
decay of excited superatoms inside the cluster. 
In Fig.~\ref{fig:HoleDynamics} we show the density of holes in 
a long lattice of superatoms. After switching on the excitation laser,
within a few lifetimes $\Gamma_s^{-1}$ of Rydberg excitations, the density 
of holes reaches an equilibrium. For large enough values of the interaction
strength $V_{\mathrm{N}} \gg w$, and thereby the laser detuning 
$(\Delta = V_{\mathrm{N}})$, the steady-state density of holes approaches the
value of $\rho_h = 1/3$.  

\begin{figure}
\label{fig:g2holes}
\centerline{\includegraphics[width=0.9\columnwidth]{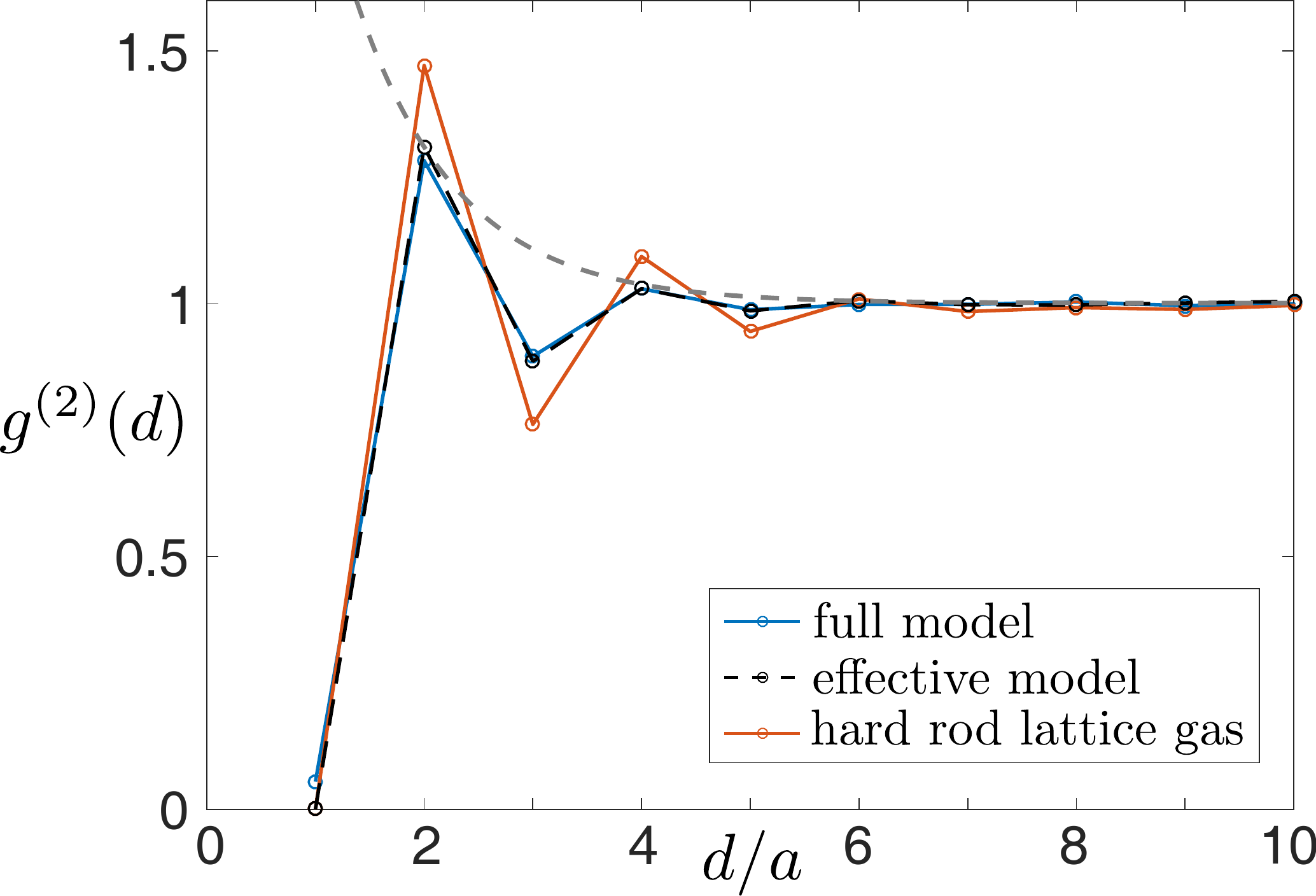}}
\caption{Second order spatial correlation function $g^{(2)}(d)$ for 
non-excited superatoms  
obtained from the numerical simulations of the full superatom model 
with $\Gamma_t/\Gamma_s \simeq 2.15$ and the effective hole model 
of Sec.~\ref{sec:HoleModel}. 
Also shown is the correlation function for the hard rod lattice gas 
with the rod length $2a$ and fixed density $1/3$ 
($\Gamma_c = \Gamma_a = 0$). 
We use an exponential fit (gray dashed line) to extract a correlation 
length of $\xi = 0.86 \pm 0.14$.}
\end{figure}

In Fig.~\ref{fig:g2holes} we show the spatial correlation function 
$g^{(2)}(d)$ for holes in the steady state. The two-particle
correlation function is defined via 
\[
g^{(2)}(d_k) \equiv \frac{ \av{\hat{n}_h^{(j)} \hat{n}_h^{(j+k)}} }
{ \av{\hat{n}_h^{(j)}}^2 }, 
\]
where $\hat{n}_h^{(j)}$ is the hole number operator for site $j$ of the 
lattice, $d_k = a k$ with $k \in \mathbb{N}$ is the distance,  
and we assume that the average hole density 
$\av{\hat{n}_h^{(j)}} = \rho_h $ is spatially uniform. 
We observe that, to a good approximation, holes behave as hard rods 
of length $2a$, with the average density of rods being $1/3$.

\subsection{Effective model for holes}
\label{sec:HoleModel}

To understand the results of the numerical simulations for the hole density 
and correlation function, we have derived an effective model for 
the equilibrium dynamics of the holes. The derivation, details 
of which are given in Appendix \ref{ap:HoleDynamicsModel}, 
is based on adiabatic elimination of short-lived configurations involving 
two or more neighboring superatoms in the ground state. Such configurations
appear when an excited superatom next to a hole decays to the ground 
state with rate $\Gamma_s$, but the lifetime of these configurations
is very short, $t \sim \Gamma_{\uparrow}^{-1}$, due to the excitation 
facilitation with the fast rate $\Gamma_{\uparrow}$. Hence, in the effective 
model, holes cannot be located on the neighboring sites, which gives the
physically intuitive picture as to why they behave as hard rods of length 
$2a$.   

There are three fundamental processes affecting the many-body dynamics 
of the holes on the lattice, as illustrated in the inset of
Fig.~\ref{fig:HoleDynamics}. 
We describe these processes in terms of the Lindblad jump operators
acting in the corresponding subspace for the holes.  

(i) Holes are created with rate $\Gamma_c = \Gamma_s$. The corresponding 
Lindblad operator is given by
\begin{equation}
\label{eq:Lcreate}
L_c^{(j)} = \sqrt{\Gamma_c} \, \sig_+^{(j)} 
[ 1- \hat{n}_h^{(j+1)} ][ 1- \hat{n}_h^{(j-1)}] ,
\end{equation}
where $\sig_\pm^{(j)}$ is the hole creation/annihilation operator 
for site $j$ of the lattice, and  
$\hat{n}_h^{(j)} \equiv \sig_+^{(j)} \sig_-^{(j)}$ is the number operator. 
The last two terms on the right hand side ensure that 
a hole cannot be created next to an existing one. 

(ii) When two holes are separated by one excitation, one of the holes 
can be annihilated in one of the following three ways described by,
\begin{subequations}
\label{eq:Lannihilate}
\begin{align}
L_{a}^{(j)} &= \sqrt{\Gamma_a/2} \, \sig_+^{(j)} \sig_-^{(j+1)} \sig_-^{(j-1)} , \\ 
L_{a_\pm}^{(j)} &= \sqrt{\Gamma_a/4} \,  \sig_-^{(j\pm 1)} \hat{n}_h^{(j)} .
\end{align}
\end{subequations}
The total annihilation probability is $\Gamma_a = \Gamma_s$. 
The remaining hole can then occupy either the middle site
with half of the total probability, or one of the side sites, each 
with quarter of the total probability.   

(iii) Holes can hop from site to site with transport rate 
$\Gamma_t = \Gamma_\mathrm{\downarrow}/2$,
\begin{equation}
\label{eq:Ltransport}
L_{t_\pm}^{(j)} = \sqrt{\Gamma_t} \, \sig_+^{(j\pm1)} \sig_-^{(j)} 
[1- \hat{n}_h^{(j\pm2)}].
\end{equation}
Here the last term ensured that the hole cannot hop to a site next
to an existing hole. 

In Fig.~\ref{fig:g2holes} we compare the spatial correlation function
$g^{(2)}(d)$ for ground state superatoms obtained from the numerical simulations
of the full superatom model and the effective hole model. We observe 
very good agreement between the full and effective models, including 
$g^{(2)}(1) \simeq 0$.

\subsection{Liquid-crystal crossover for lattice holes}
\label{sec:extended-model}

Although in the present setup the hole hopping rate cannot be made
arbitrary small, $\Gamma_t \gtrsim \Gamma_{c,a}$, it is instructive
to analyze how varying $\Gamma_t$ would affect the many-body steady state.

Consider first the hypothetical case of no hole transport, $\Gamma_t \to 0$. 
We can also neglect for now the probability of refilling the hole, due to 
smallness of the corresponding rate $\Gamma_\mathrm{seed}$. Then the
only stable configuration corresponds to holes on every third site
of the lattice, since neither the hole creation nor annihilation
processes of Eqs.~(\ref{eq:Lcreate}) and (\ref{eq:Lannihilate}) 
affect the system. Due to translational invariance of the lattice, 
this configuration is triple degenerate.
The hole density-density correlation function $g^{(2)}(d)$ is peaked 
at $d=3a$, and the steady state approaches a pure state with long-range 
crystalline order. 
Since the density of holes is exactly $\rho_h =1/3$, their number does 
not fluctuate. We can characterize the number fluctuations of the holes 
by the Mandel $Q$ parameter
\[
Q \equiv \frac{\av{\hat{n}_h^2} - \av{\hat{n}_h}^2}{\av{\hat{n}_h}} - 1 ,
\]
where $\hat{n}_h \equiv \sum_j \hat{n}_h^{(j)}$ is the total number of holes.
$Q < 0$ signifies sub-Poissonian number distribution, with $Q = -1$ 
corresponding to a precise number of holes with no fluctuations. 

\begin{figure}
\label{fig:avrhoQg223}
\centerline{\includegraphics[width=1\columnwidth]{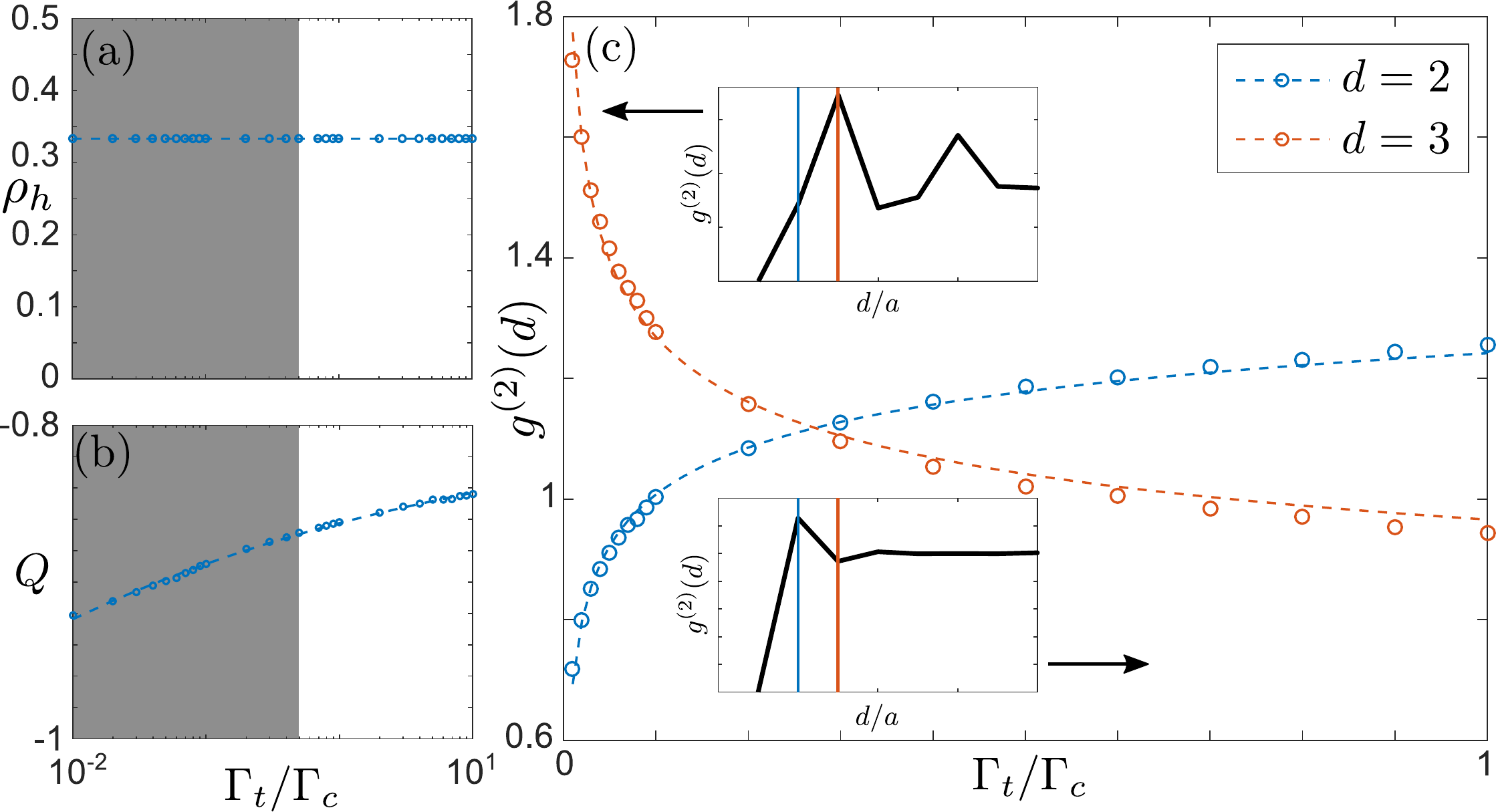}}
\caption{(a) Average density of holes $\rho_h$, 
(b) Mandel $Q$ parameter for the total number of holes, and 
(c) amplitudes of the density-density correlation function $g^{(2)}(d)$ 
for the periods of $d=2a$ (blue circles) 
and $d=3a$ (red circles), versus the hopping rate $\Gamma_t$.  
Dashed lines indicate an algebraic fit function. 
The shaded areas in (a) and (b) indicate the hypothetical 
regime of $\Gamma_t < \hlf \Gamma_s$ not accessible in the current model. }
\end{figure}

As we now increase $\Gamma_t$, the holes become mobile and the peak
of the correlation function $g^{(2)}(d)$ at $d=3a$ is progressively
reduced, see Fig.~\ref{fig:avrhoQg223}. The holes can now approach
each other and annihilate, followed by hole creation on the allowed 
sites, which causes finite fluctuation of the hole number. 
Yet, their statistics remains highly sub-Poissonian,
$Q \simeq -0.81$, even for large $\Gamma_t \gg \Gamma_{c,a}$. 
Since the hole creation and annihilation rates are the same, 
$\Gamma_c = \Gamma_a = \Gamma_s$, their mean density stays 
close to $\rho_h = 1/3$. Interestingly, the crystal with periodicity 
$d=3a$ does not simply melt into a liquid with the same period. 
Rather, for a large hopping rate, the correlation 
function $g^{(2)}(d)$ exhibit period $d=2a$ with short correlation 
length $\xi \lesssim 1a$, see Fig.~\ref{fig:g2holes}. Since holes 
cannot come closer than two lattice sites, they start to behave 
as mobile hard rods of length $2a$.  

We finally note that when the interaction strength $V_{\mathrm{N}}$,
and thereby the laser detuning $\Delta = V_{\mathrm{N}}$, are not sufficiently 
larger than the excitation linewidth, the seed rate $\Gamma_{\mathrm{seed}}$ 
is not negligible and holes can refill, leading to $\rho_h < 1/3$, 
as can be seen in Fig.~\ref{fig:HoleDynamics}. 

\subsection{Transient excitation dynamics of the system}

\begin{figure}
\label{fig:ExcitationDynamics}
\centerline{\includegraphics[width=1.0\columnwidth]{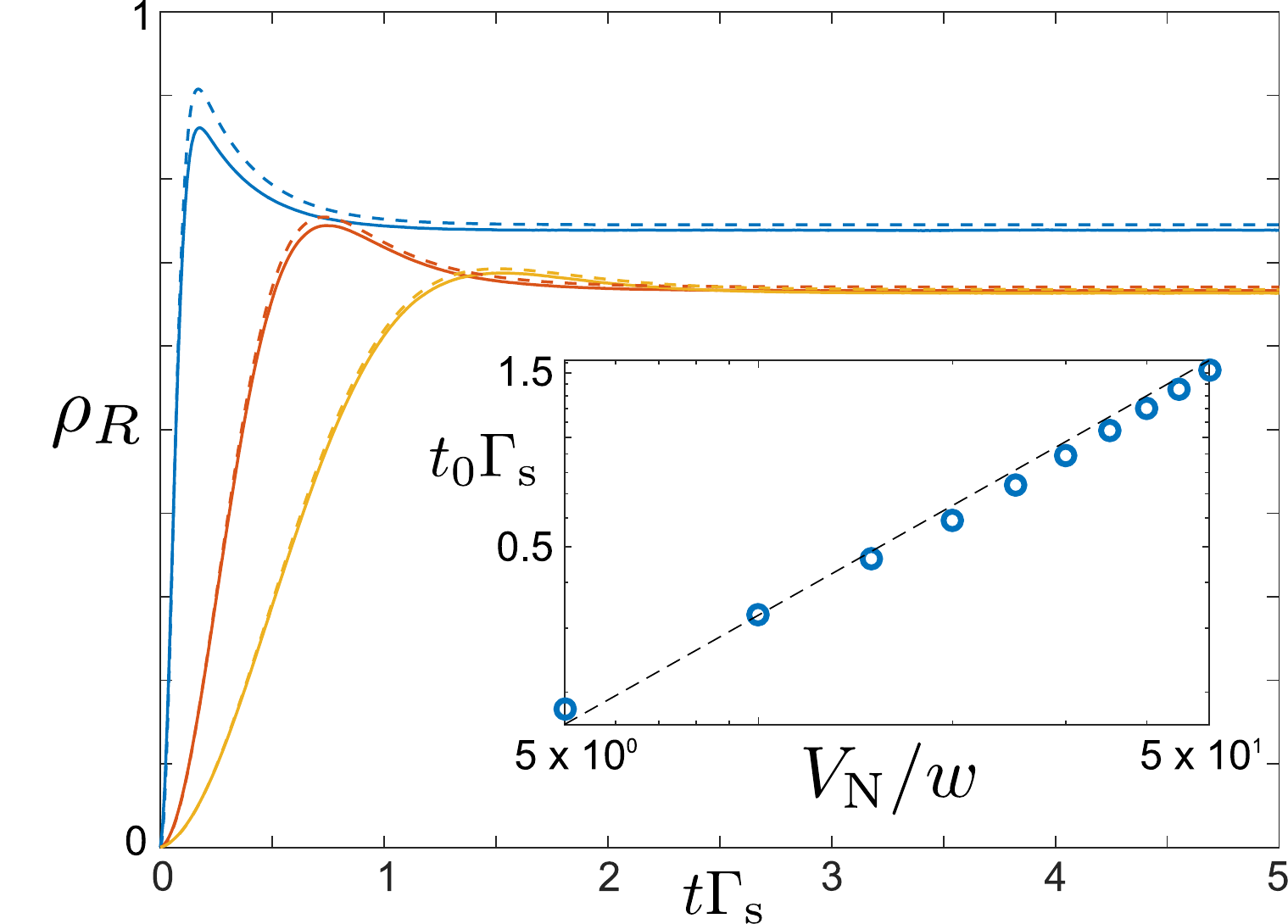}}
\caption{Excitation dynamics of the lattice of superatoms, 
initially all in the ground state, as obtained from the simulations 
of the microscopic rate equations (solid lines) and macroscopic model 
(dashed lines). The parameters and color code are the same as in 
Fig.~\ref{fig:HoleDynamics} and the initial state has one seed excitation. 
Inset: Scaling of the peak time $t_0$ with the interaction strength 
$V_\mathrm{N}/\omega$. The dashed line corresponds to the analytic 
estimate $t_0 \simeq 2/\sqrt{\Gamma_\mathrm{seed}\Gamma_\uparrow}$.}
\end{figure}

As promised above, we now discuss the dynamics of the system of superatoms 
initially in the ground state upon switching on the excitation laser. 
In Fig.~\ref{fig:ExcitationDynamics} we show the density $\rho_{\mathrm{R}}$ 
of Rydberg excitations as a function of time, for different values 
of the interaction strength $V_{\mathrm{N}}$ (and laser detuning 
$\Delta = V_{\mathrm{N}}$). In the long time limit $t \gg \Gamma_s^{-1}$, 
for large enough $V_{\mathrm{N}} \gg w$, the system reaches the steady state 
with the Rydberg excitation density $\rho_{\mathrm{R}} \simeq 2/3$, 
consistent with the hole density $\rho_h \simeq 1/3$ analyzed above.
For smaller values of $V_{\mathrm{N}}$, we have in the steady state 
$\rho_{\mathrm{R}} > 2/3$, since holes can be refilled with 
appreciable seed rate $\Gamma_{\mathrm{seed}}$.

Consider now the transient regime $0<t \lesssim \Gamma_s^{-1}$. 
As seen in Fig.~\ref{fig:ExcitationDynamics}, after switching 
on the laser, the density of Rydberg excitations can peak at a 
large value and then relax to the lower steady-state density. 
At first sight, this observation is surprising, since such a behavior 
is reminiscent to partially coherent dynamics of a quantum system, 
such as, e.g., damped Rabi oscillations, while here our system is 
completely governed by rate equations and no coherences are involved.
We now outline a macroscopic model that will explain the nature of
the peak and the associated peak time $t_0$. 

We consider three basic states of the system: The ground state (g), 
the fully excited state (e) and the final steady state (s). The
corresponding probabilities are denoted by $p_{\mathrm{g}}$,
$p_{\mathrm{e}}$, and $p_{\mathrm{s}}$. Initially the system is in 
the ground state, $p_{\mathrm{g}} = 1$, and to start the dynamics 
we need a seed excitation. The probability of the seed $p_\mathrm{seed}$ 
is governed by the equation 
$\dt p_\mathrm{seed}(t) = p_{\mathrm{g}} \Gamma_\mathrm{seed}$.
For short times, we can assume $p_{\mathrm{g}} \simeq 1$ and obtain 
linear growth of the seed probability, $p_\mathrm{seed}(t) 
\simeq p_\mathrm{seed}(0) + \Gamma_\mathrm{seed} t$.
For longer times, this approximation breaks down, but once 
$p_{\mathrm{g}}$ is depleted, the role of the seed becomes unimportant,
as will become clear shortly.

The equations of motion for the probabilities of the three basic states are
\begin{subequations}
\begin{align}
\dt p_{\mathrm{g}} &= - p_\mathrm{seed} \, 2 \Gamma_\uparrow \, p_{\mathrm{g}} , \\
\dt p_{\mathrm{e}} &= + p_\mathrm{seed} \, 2 \Gamma_\uparrow \, p_{\mathrm{g}} 
- 3\Gamma_s \, p_{\mathrm{s}} + \Gamma_\mathrm{seed} \, p_{\mathrm{s}} ,\\
\dt p_{\mathrm{s}} &= + 3\Gamma_s \, p_{\mathrm{s}} 
- \Gamma_\mathrm{seed} \, p_{\mathrm{s}}.
\end{align}
\end{subequations}
Each seed excitation in the lattice triggers fast growth of 
facilitated excitations in both lattice directions. The ground state is then 
being depleted with the rate $2 \Gamma_\uparrow$ and its population is
transferred into the fully excited state. In turn, the Rydberg excitations 
in the fully excited state decay with rate $\Gamma_s$. Since the steady state
corresponds to configurations with, on average, every third site 
non-excited, the decay rate of $p_{\mathrm{e}}$ to $p_{\mathrm{s}}$ is $3\Gamma_s$,
which we have also verified via numerical simulations of the full system
with all the superatoms initially excited. We also include in the 
above equations the process of refilling the holes with the corresponding
rate $\Gamma_\mathrm{seed}$. Then the solution of these equations 
approximates remarkably well the mean density of Rydberg excitations
$\rho_{\mathrm{R}}(t) = p_{\mathrm{e}}(t) + \tth p_{\mathrm{s}}(t)$, even 
in the regime of sizable $\Gamma_\mathrm{seed}$, as can be seen in the
main panel of Fig.~\ref{fig:ExcitationDynamics}. 

The peak in the excitation density $\rho_{\mathrm{R}}$ is reached when 
the ground state probability $p_{\mathrm{g}}$ is depleted and the majority
of superatoms are transferred to the excited state, $p_{\mathrm{e}} \simeq 1$.
Integration of $\dt p_{\mathrm{g}}$ suggests the peak time scaling as
$t_0 \propto 1/\sqrt{\Gamma_\mathrm{seed}\Gamma_\uparrow} \propto V_\mathrm{N}$.
The scaling of the peak time with the interaction strength $V_\mathrm{N}/w$
is verified in the inset of Fig. \ref{fig:ExcitationDynamics}.
Note finally that our macrocsopic model neglects cluster collisions, 
also producing holes, and thereby slightly overestimates the excitation 
density $\rho_R$ obtained from the simulations of the complete microscopic
model.

\section{Conclusions and outlook}
\label{sec:Conclusion}

To summarize, we have studied the excitation dynamics and steady state of 
a lattice of Rydberg superatoms driven by a laser in the facilitation regime. 
We have shown that the steady state of the system has nearly universal
Rydberg excitation density of $\rho_\mathrm{R} = 2/3$. More interestingly, 
it corresponds to an equilibrium dynamics of mobile quasi-particles 
-- excitation holes. We have derived an effective hole model which
involves hole creation and pair annihilation of holes separated by
two lattice sites. We have found that the number fluctuations of the 
holes are characterized by the Mandel $Q$ parameter $Q \simeq -0.81$, 
and their spatial correlations decay on a distance of $\xi \lesssim 1a$
comparable to the lattice constant $a$.

That negative values of $Q$, which signify pronounced sub-Poissonian 
number distribution, have their origin in the hard rod constraint, 
has been pointed out already in \cite{Ates2012a} considering a Gibbs 
ensemble of Rydberg excitations. In our system, however, the pair 
annihilation of holes leads to a state with much stronger suppressed 
fluctuations. In the Gibbs state of Ref. \cite{Ates2012a}, we would 
have to choose a rod length of about $d= 0.611$ to obtain the average 
hole density $\rho_h = 1/3$, but this would then lead to $Q \simeq -0.63$ 
which is significantly larger than what we obtain from our simulations.

Our model corresponds to an experimentally amenable regime with a large
hole transport rate $\Gamma_t \sim \Gamma_{c,a}$ comparable to the hole
creation and annihilation rates $\Gamma_{c,a} = \Gamma_s$, which, in turn, 
are determined by the spontaneous decay rate of Rydberg excitations.
The resulting spatial correlations of the holes have a period of 
$d=2a$ and short correlation length $\xi \simeq a$. If we freeze the hole 
motion, $\Gamma_t \ll \Gamma_s$, we would obtain long-range order,
$\xi \gg a$, with $d = 3a$ periodicity. Such a regime can in principle
be achieved, but at the expense of more complicated experimental setup 
involving additional lasers coupling the ground state of superatoms 
to a different long lived Rydberg or metastable state which would 
make the holes immobile.

Another interesting direction of research is to consider different lattice 
geometries, e.g., a two-dimensional square or triangular lattice. The latter 
might simulate dissipative frustrated spin models. Finally, the rate equation 
approach, amenable to large scale numerical simulations, is applicable in 
the regime of strong dephasing. Coherence effects might lead to interesting 
dynamics and yield long range correlations and entanglement. However, 
fully quantum many-body simulations are limited to small system sizes.

\begin{acknowledgments}
This work was supported by DFG through SFB/TR185 (M.F. and F.L.) 
and the H2020 FET Proactive project RySQ (D.P.).
F.L. is supported by a fellowship through the Excellence Initiative MAINZ 
(DFG/GSC 266).
D.P. is grateful to the University of Kaiserslautern for hospitality and to 
the Alexander von Humboldt Foundation for support during his stay in Germany.
\end{acknowledgments}

\appendix

\section{Rate equations model for a superatom}
\label{ap:SAREM}

Here we consider in some detail the dissipative dynamics of a single Rydberg 
superatom. The superatom consists of $N$ two-level atoms within a volume 
of linear dimension $\Delta r$ smaller than the Rydberg blockade distance 
$a_{\mathrm{B}}$. We define $a_{\mathrm{B}}$ as the distance below which 
the interatomic interaction strength $V(r \leq a_{\mathrm{B}}) \geq w$ 
starts to exceed the steady-state excitation linewidth of an atom,  
$w = \sqrt{4 \Omega^2 \gamma/\Gamma_s + \gamma^2} 
\simeq 2 \Omega \sqrt{\gamma/\Gamma_s}$ \cite{Petrosyan2013,Petrosyan2013c}. 
Here $\gamma \equiv \frac{1}{2}(\Gamma_d + \Gamma_s)$, and we assume 
$\Omega^2 \gg \Gamma_s \Gamma_d$.
Due to the strong Rydberg blockade, the superatom can accommodate at most 
one Rydberg excitation. 

The collective ground state of a superatom 
$\ket{G} \equiv \ket{g_{1} g_2 \ldots g_{N}}$ is coherently 
coupled by the laser to the symmetric single excitation state 
$\ket{E_s} \equiv \frac{1}{\sqrt{N}} 
\sum_k \ket{g_1 g_2 \ldots e_k \ldots g_{N}}$ 
with Rabi frequency $\sqrt{N} \Omega$,
see Fig.~\ref{fig:SuperatomLevelScheme}(a). In addition, there are 
$N - 1$ non-symmetric single excitation states $\{ \ket{E_{ns}}_m \}$
which are not directly coupled to the ground state by the laser. 
All the excited states $\{ \ket{E} \}$ spontaneously decay to the ground 
state $\ket{G}$ with rate $\Gamma_s$. The dephasing $\Gamma_d$ of the atomic 
Rydberg state $\ket{e}$, with respect to the ground state $\ket{g}$,  
leads to incoherent coupling of any single excitation state $\ket{E}$ 
of the superatom to any other such state $\ket{E'}$ with rate $\Gamma_d/N$. 
We may replace the manifold $\{ \ket{E_{ns}}_m \}$ with a single aggregate 
non-symmetric state $\ket{E_{ns}}$, obtaining for the superatom an 
effective three-level system, $\{\ket{G}, \ket{E_s}, \ket{E_{ns}}\}$.
We describe this dissipative system with a ``vector'' of density matrix elements
$\vec{\rho}=(\rho_{GG}, \rho_{E_sE_s},\rho_{GE_s}, \rho_{E_s G},\rho_{E_{ns} E_{ns}})^T$
which obeys the equation of motion
\begin{equation}
\label{eq:SA-EOM}
\partial_t \vec{\rho} = \Lambda \vec{\rho},
\end{equation}
where 
\begin{widetext}
\begin{equation}
\label{eq:Lambda}
\Lambda = 
\begin{pmatrix}
0 & \Gamma_s & i \sqrt{N} \Omega & -i \sqrt{N} \Omega & \Gamma_s  \\

0 & -\Gamma_s-\frac{N-1}{N}\Gamma_d & -i \sqrt{N} \Omega & i \sqrt{N} \Omega & \frac{1}{N} \Gamma_d \\

i \sqrt{N} \Omega & - i \sqrt{N} \Omega & - i \Delta - \gamma & 0 & 0 \\

-i \sqrt{N} \Omega & i \sqrt{N} \Omega & 0 & i \Delta - \gamma & 0 \\

0 & \frac{N-1}{N}\Gamma_d & 0 & 0 & -\Gamma_s-\frac{1}{N} \Gamma_d
\end{pmatrix}.
\end{equation}
\end{widetext}
For $N = 1$, we retrieve the well known optical Bloch equations for 
a single two level atom, with the decay rate $\Gamma_s$ of the excited 
state population $\rho_{ee}$ and the relaxation rate $\gamma 
= \frac{1}{2}(\Gamma_s+\Gamma_d)$ of the coherence $\rho_{ge}$. For 
a superatom with $N > 1$, dephasing induces population $\rho_{E_{ns} E_{ns}} >0$
of non-symmetric state(s), but their coherences remain decoupled, 
$\rho_{E_{ns} G} = \rho_{E_{ns} E_s} =0$. By definition, the superatom 
can contain at most one Rydberg excitation, 
$\rho_{GG} + \rho_{E_s E_s} + \rho_{E_{ns} E_{ns} } = 1$.

Solving Eq.~(\ref{eq:SA-EOM}) in the steady state, $\partial_t \vec{\rho} =0$,
we obtain the following expression for the total excited state population 
of the superatom, 
\begin{multline}
\label{eq:rhoEE}
\rho_{EE} \equiv \rho_{E_s E_s} + \rho_{E_{ns} E_{ns} } 
\\ = \frac{ 2 N \Omega^2 \gamma}
{ \Omega^2 ( N + 1 ) 
\left( \Gamma_d + \frac{2 N}{ N + 1 } \Gamma_s \right) 
+ \Gamma_s ( \gamma^2 + \Delta^2 )}.
\end{multline}
For $N=1$, this reduces to the excited state population 
of a single two level atom \cite{Petrosyan2013},
\begin{equation}
\rho_{ee} = \frac{ \Omega^2 }
{ 2 \Omega^2 + \frac{\Gamma_s}{2 \gamma} ( \gamma^2 + \Delta^2 )},
\label{eq:rhoee}
\end{equation}
which is bounded by $\rho_{ee} \leq 0.5$ even for resonant excitation 
$|\Delta| \ll \gamma$.
On the other hand, for large $N$ and small decay rate
$\Gamma_s \ll \Gamma_d, \Omega^2/\Gamma_d$, the excited state population 
of the superatom is approximately given by 
\begin{equation}
\rho_{EE} \simeq  \frac{N}{N + 1},
\end{equation}
which quickly approaches $\rho_{EE} \simeq 1$ with increasing $N$. 
We emphasize that the steady-state population inversion of the superatom
is brought about by strong driving and dephasing, which tend to equalize 
populations of all $N + 1$ states, i.e.,  $N$ single excitation 
states and the ground state. In contrast, without dephasing 
($\Gamma_d \ll \Gamma_s$), the superatom reduces to a two-level system, 
with the ground state $\ket{G}$ coherently coupled to the symmetric 
excited state $\ket{E_s}$ which decays back to the ground state with 
rate $\Gamma_s$, and the resulting steady-state Rydberg excitation 
probability is $\rho_{EE} \simeq \rho_{E_s E_s} \leq 1/2$.

From Eq.~(\ref{eq:rhoEE}), with strong dephasing $\Gamma_d \gg \Gamma_s$ 
and $N \gg 1$, we obtain the excitation linewidth 
of the superatom 
\begin{equation}
w_\mathrm{SA} \simeq \sqrt{2 N \Omega^2 \gamma / \Gamma_s+\gamma^2}.
\end{equation}
Comparing it with the single-atom excitation linewidth $w$ which follows from
Eq.~(\ref{eq:rhoee}), we notice the analogy upon replacement 
$\Omega^2 \to \frac{1}{2} N\Omega^2$, rather than 
$\Omega^2 \to N\Omega^2$ as one would naively expect from
the collective enhancement of the $N$-atom Rabi frequency 
$\sqrt{N} \Omega$. This factor of $2$ difference stems from 
the fact that the Rydberg excitation probability of a saturated superatom
approaches unity, rather than $\frac{1}{2}$.

For strong dephasing $\Gamma_d \gtrsim \Omega$, the dynamics of a superatom can
be described, to a good approximation, by rate equations which we now derive.
In the equations for the density matrix elements, 
we adiabatically eliminate the coherences 
$\rho_{E_s G}$, $\rho_{G E_s}$, obtaining rate equations 
$\partial_t \vec{\rho} = \Lambda_3 \vec{\rho}$ for populations
$\vec{\rho}=(\rho_{GG}, \rho_{E_sE_s},\rho_{E_{ns} E_{ns}})^T$ with
\begin{equation}
\label{eq:Lambdap3}
\Lambda_3 = 
\begin{pmatrix}
-N \chi & N \chi +\Gamma_s & \Gamma_s  \\
N \chi & - N \chi - \Gamma_s - \frac{N-1}{N}\Gamma_d & \frac{1}{N} \Gamma_d \\
0 & \frac{N-1}{N}\Gamma_d & -\Gamma_s - \frac{1}{N} \Gamma_d
\end{pmatrix} ,
\end{equation}
where 
\[
\chi \equiv \frac{ 2 \Omega^2 \gamma}{\gamma^2 + \Delta^2} .
\] 
The laser tends to equalize populations $\rho_{GG}$ and $\rho_{E_sE_s}$ 
of the ground and symmetric excited states. In turn, strong dephasing quickly 
transfers the population of the symmetric state to the non-symmetric states.
For large $N$, the reverse transfer from $\rho_{E_{ns} E_{ns}}$ to $\rho_{E_s E_s}$
is suppressed by a factor of $1/N$. The symmetric state then plays 
the role of an intermediate state having small population 
$\rho_{E_sE_s} \sim \rho_{E_{ns} E_{ns}}/N$. 
Upon adiabatic elimination of $\rho_{E_sE_s}$ we finally obtain rate equations 
$\partial_t \vec{\rho} = \Lambda_2 \vec{\rho}$ for the populations
$\vec{\rho}=(\rho_{GG}, \rho_{EE})^T$ of the ground and excited states
($\rho_{GG} + \rho_{EE} \simeq 1$), 
\begin{equation}
\label{eq:Lambdap2}
\Lambda_2 = 
\begin{pmatrix}
- \Gamma_\mathrm{ex} & \Gamma_\mathrm{de}  \\
\Gamma_\mathrm{ex} & - \Gamma_\mathrm{de}
\end{pmatrix},
\end{equation}
where the excitation and de-excitation rates are given by 
\begin{align}
\label{eq:Gammaexdeapp}
\Gamma_\mathrm{ex} &= \frac{(N-1)\chi}{N \chi + 2 \gamma} \Gamma_d, \\
\Gamma_\mathrm{de} &= \frac{\chi}{N \chi + 2 \gamma} \Gamma_d + \Gamma_s.
\end{align}
This is illustrated in Fig. \ref{fig:SuperatomLevelScheme}(b).

\begin{figure}[t]
  \centerline{\includegraphics[width=1 \columnwidth]{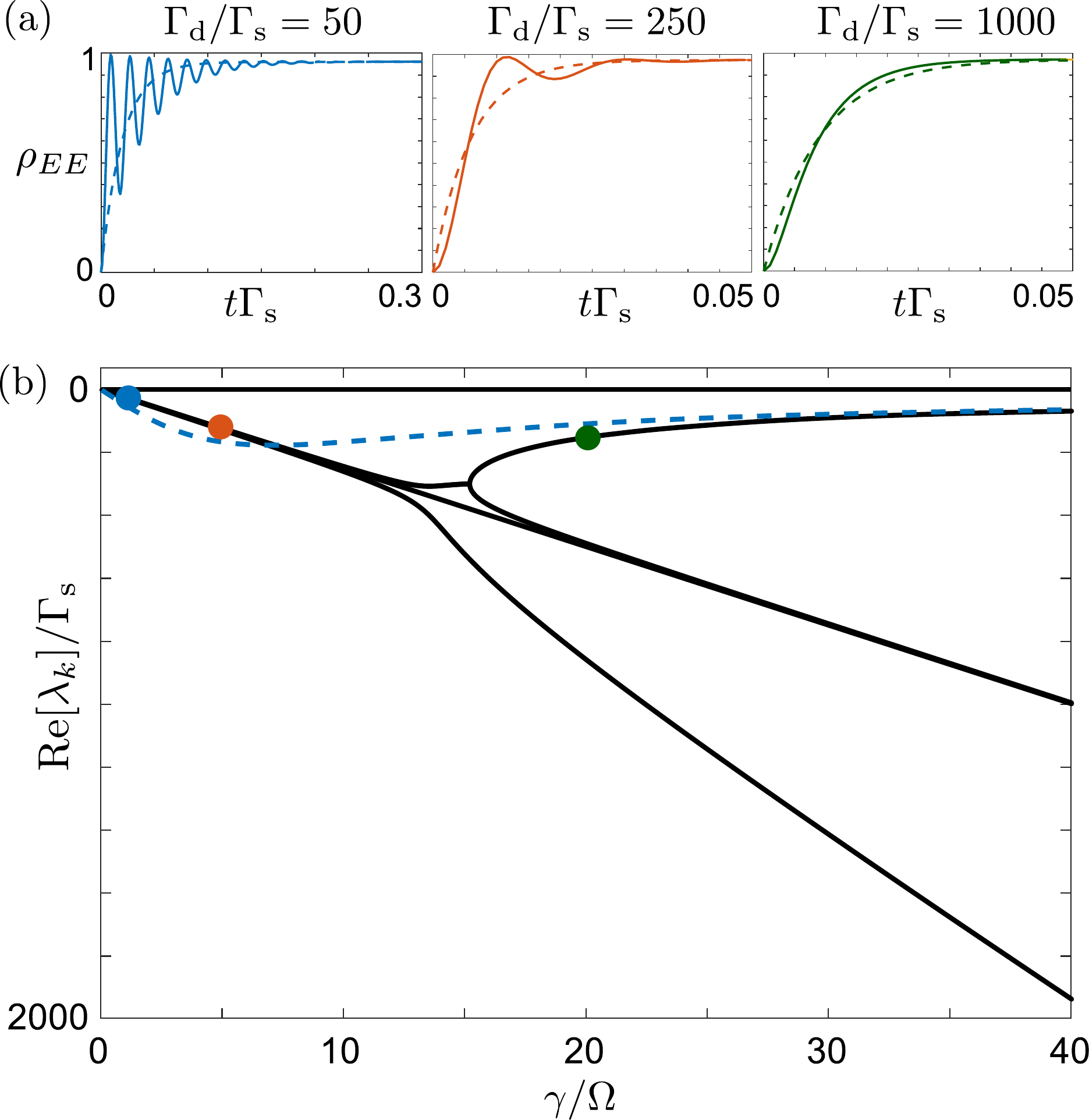}}
  \caption{(a) Excitation dynamics of a single superatom, 
    containing $N = 50$ atoms, as obtained from 
    the rate equation model (dashed lines), and 
    the exact solution of the density matrix equations in the single 
    excitation Hilbert space (solid lines). 
    The laser is resonant $\Delta =0$ and has the single-atom Rabi frequency 
    $\Omega = 25 \Gamma_s$, while the dephasing rate is 
    $\Gamma_d/\Gamma_s = 50,250,1000$ in the left, middle, right graphs, 
    respectively. 
    (b) Real part of eigenvalues $\lambda_k$ of $\Lambda$ in 
    Eq.~\eqref{eq:Lambda}, versus the transverse atomic relaxation rate 
    $\gamma$. The (blue) dashed line shows the total relaxation rate 
    of a superatom $\Gamma_{\mathrm{tot}} = \Gamma_\mathrm{ex}+\Gamma_\mathrm{de}$ 
    obtained from the rate equation model.
  } 
\label{appfig:SADynLiuvSp}
\end{figure}

In Fig. \ref{appfig:SADynLiuvSp}(a) we compare the excitation dynamics of a  
resonantly driven superatom as obtained from the solution of the rate equations
and the exact solution of the master equation for the density operator using 
the Monte Carlo wavefunction approach in the truncated single excitation 
Hilbert space. We observe that the rate equations provide accurate description
of the dynamics of the system when $\gamma \gtrsim \sqrt{N}\Omega$, while
they cannot account for the (damped) Rabi oscillations between the ground and 
the excited states when $\sqrt{N}\Omega \gg \gamma$. Nevertheless, the 
rate equations model approximates well the relaxation timescale and the 
steady state population of the superatom.

In Fig. \ref{appfig:SADynLiuvSp}(b) we show the spectrum of eigenvalues 
$\lambda_k$ of the matrix $\Lambda$ in Eq.~(\ref{eq:Lambda}). The eigenvalue
$\lambda_0 = 0$ corresponds to the steady state of the system, while 
the (negative) real parts of the other eigenvalues, $\mathrm{Re} [\lambda_k]$ 
for $k=1,2,3,4$, characterize the relaxation rates of the superatom towards 
the steady state. In that figure, we also show the total relaxation rate 
of a superatom towards the steady state as given by the rate equations model, 
$\Gamma_{\mathrm{tot}} = \Gamma_\mathrm{ex}+\Gamma_\mathrm{de}$, which compares 
favorably with the exact relaxation rate for a wide range of parameters.

%

\section{Excitation facilitation conditions for a chain of superatoms}
\label{ap:AntiBlockadeConditions}

With the interatomic potential $V(r) = C_p/r^p$ (assuming 
repulsive interaction $C_p>0$) and the single atom excitation 
linewidth $w$,  the Rydberg blockade distance is defined as
$a_\mathrm{B} \equiv (C_p/w)^{1/p}$. Our model assumes that each superatom 
can accommodate at most one Rydberg excitation. We therefore require
the spatial extent of a superatom to be small compared to blockade 
distance, $\Delta r \ll a_\mathrm{B}$.

Clearly, the superatom excitation probability is maximal at resonant driving 
and is suppressed when the laser is detuned by more than its excitation
linewidth $w_\mathrm{SA}$. We consider a one-dimensional chain of superatoms 
driven by spatially uniform laser with detuning $\Delta$. We set the laser 
detuning to be equal to the interaction strength between neighboring 
superatoms separated by the lattice constant $a$, $\Delta = V(a)$. 
Then an already excited superatom will facilitate the excitation of 
the nearest neighbors by shifting the Rydberg energy level into 
resonance with the laser field. On the other hand, we require 
that $\Delta > w_\mathrm{SA}$ in order to suppress excitation of superatoms 
that have either two non-excited neighbors or two excited neighbors. 
Since $w_\mathrm{SA} > w$, the facilitation distance 
$r_\mathrm{fac} = (C_p/\Delta )^{1/p}$, and thereby the lattice constant 
$a = r_\mathrm{fac}$, are smaller than the blockade distance $a_\mathrm{B}$.

The interaction potential $V(r)$ is a convex monotonic function.
Our approach is valid when the interaction induced Rydberg level 
shifts of all the atoms of the facilitated superatom are within 
the superatom linewidth $w_\mathrm{SA}$. Linearizing the interaction 
potential around the facilitation distance $r_\mathrm{fac}$, we obtain the 
following condition on the spatial extent $\Delta r$ of the superatom
\begin{equation}
\Delta r \lesssim \frac{1}{p} \frac{r_\mathrm{fac}^{p+1}}{C_p/ w_\SA} 
 = \frac{1}{p} \frac{w_\mathrm{SA}}{\Delta} a .
\end{equation}

Finally, in our treatment of the 1D lattice of superatoms, we neglect the 
interaction shifts of the next neighbors, $V(2 a) \lesssim w_\mathrm{SA}$. 
This can be rewritten as a condition on the power-law scaling $p$ 
of the interaction potential,
\begin{equation}
p \log(2) \geq \log(\Delta/w_\mathrm{SA}) .
\end{equation}

\section{Derivation of the effective dynamical model for holes}
\label{ap:HoleDynamicsModel}

\begin{figure}
\label{fig:EffectiveHoleModel}
\centerline{\includegraphics[width=0.8\columnwidth]{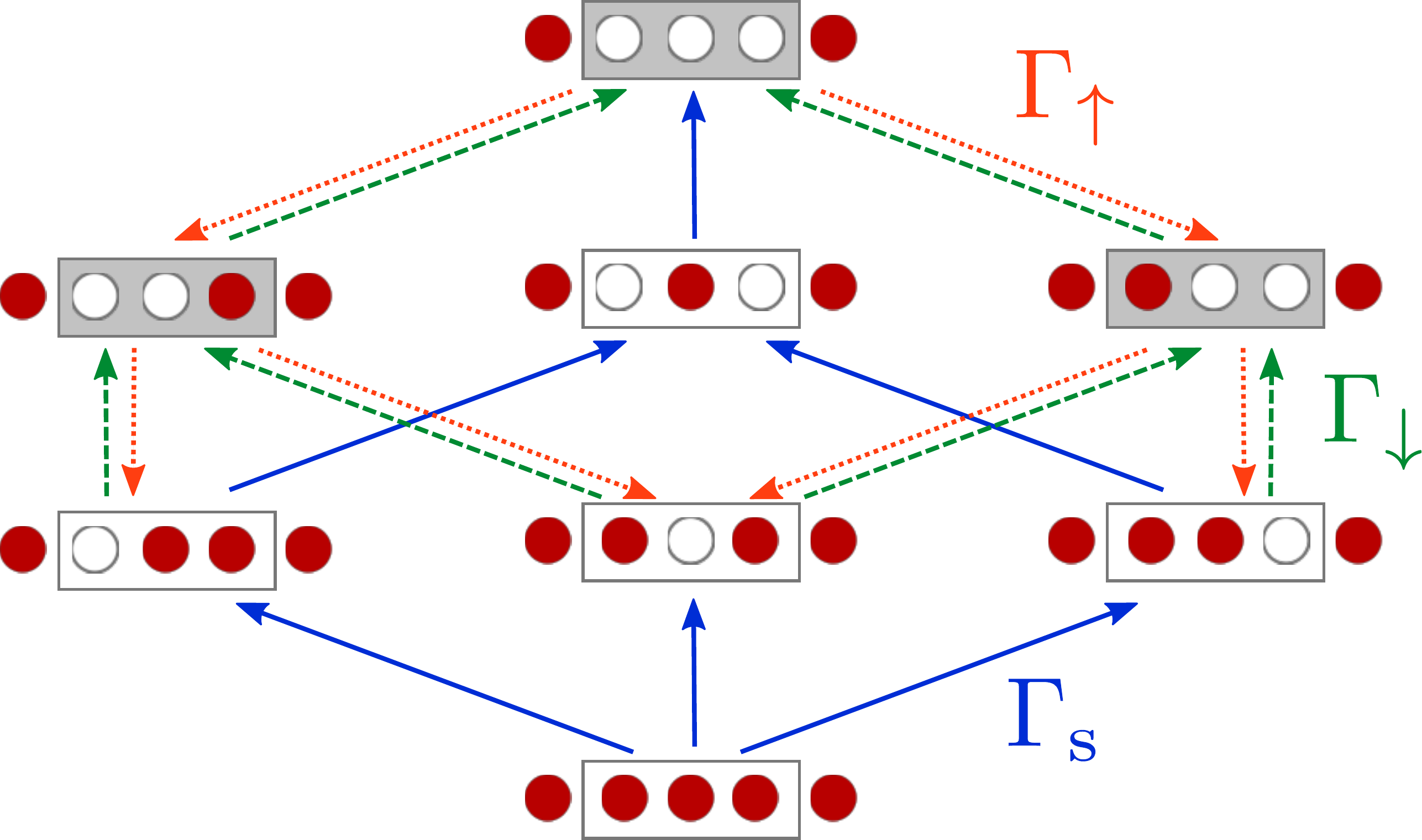}}
\vspace{0.6cm}
\centerline{\includegraphics[width=0.8\columnwidth]{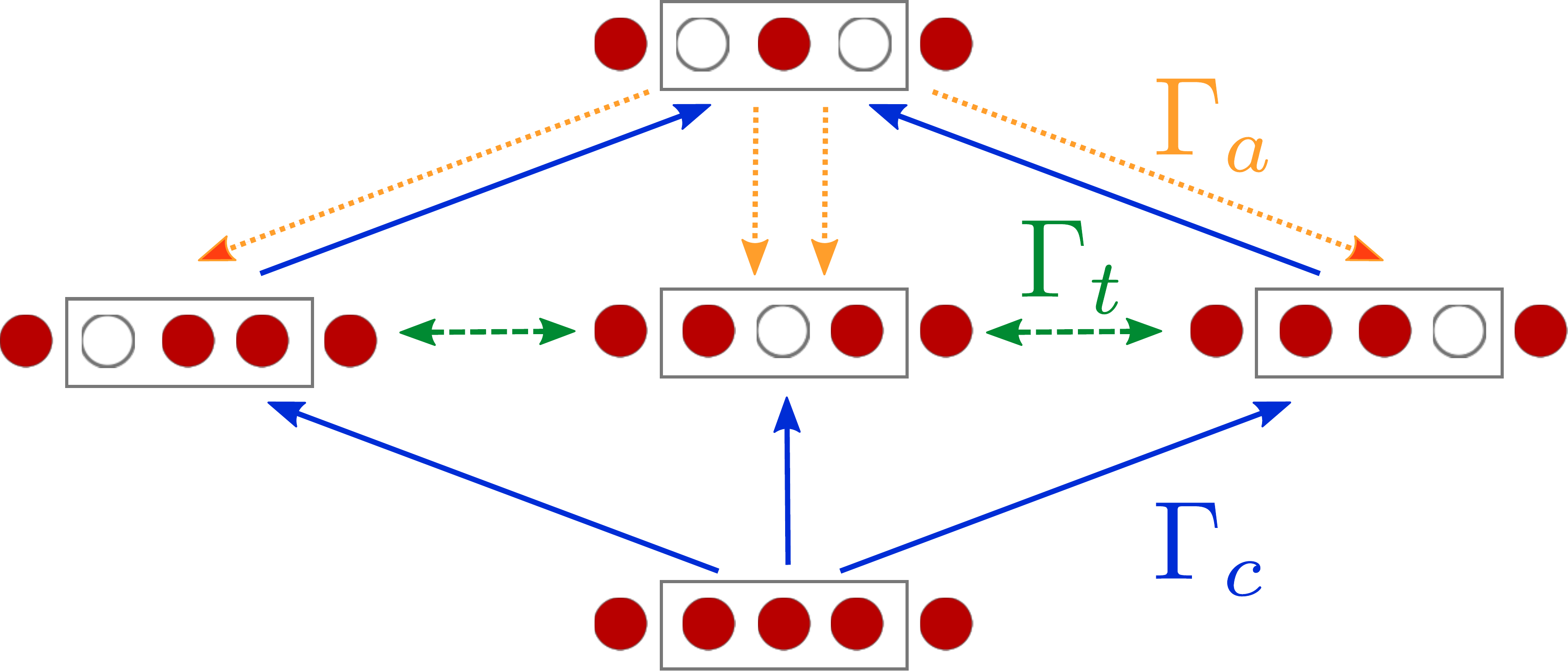}}
\caption{Sketch for the derivation of the effective hole model. 
Filled (red) circles denote excited superatoms,
and open (white) circles the non-excited ones.  
Top: Using a three-site chain of superatoms, and fixed excited superatoms 
on the left and right of the chain, we identify the transition rates 
$\Gamma_s, \Gamma_{\downarrow}, \Gamma_{\uparrow}$ between the various excitation 
configurations. All configurations with two or more neighboring non-excited
superatoms are adiabatically eliminated (gray shaded configurations).
Bottom: The resulting effective model for hole dynamics involves three 
processes: creation of holes with rate $\Gamma_c$, annihilation of 
holes with rate $\Gamma_a$, and transport of holes with rate $\Gamma_t$.}
\end{figure}

Consider a small chain of superatoms shown in Fig.~\ref{fig:EffectiveHoleModel}.
We focus on a subset of the system containing three superatoms that can be
in the ground state or excited to the Rydberg state. We denote the probability 
of each configuration $\{a,b,c\}$ by $p_{abc}$, where $a,b,c = 0, 1$ 
for non-excited (0) and excited (1) superatom on the corresponding site.
Starting with all the superatoms excited, $\{1,1,1\}$, we can create a hole on 
any one site, e.g., $\{0,1,1\}$, with the spontaneous decay rate $\Gamma_s$. 
The reverse process of exciting a hole can be neglected due to smallness of 
the transition rate $\Gamma_\mathrm{seed} \ll \Gamma_{\uparrow,\downarrow}, \Gamma_s$. 
Next, an excited superatom neighboring a non-excited one can be de-excited,
e.g., $\{0,1,1\} \to \{0,0,1\}$, with rate $\Gamma_{\downarrow}$. 
An excited superatom between two non-excited ones can be de-excited,
e.g., $\{0,1,0\} \to \{0,0,0\}$, with the spontaneous decay rate 
$\Gamma_s$. Finally, the probability of configuration with two 
or more neighboring non-excited superatoms will quickly decay, 
e.g., $\{0,0,1\} \to \{1,0,1\}$ or $\{0,1,1\}$, with the the facilitated 
excitation rate $\Gamma_{\uparrow}$. We thus obtain the following set of
equations for the probabilities of various configurations 
(see Fig.~\ref{fig:EffectiveHoleModel} upper panel),
\begin{align}
\partial_t p_{111} =& -3 \Gamma_s p_{111} , \\
\partial_t p_{011} =& - (\Gamma_s + \Gamma_\downarrow) p_{011} + \Gamma_s p_{111}
+ \Gamma_\uparrow p_{001} , \\
\partial_t p_{101} =& - 2\Gamma_\downarrow p_{101} + \Gamma_s p_{111}
+ \Gamma_\uparrow p_{001} 
 + \Gamma_\uparrow p_{100} , \\
\partial_t p_{110} =& - (\Gamma_s + \Gamma_\downarrow) p_{110} 
+ \Gamma_s p_{111} + \Gamma_\uparrow p_{100} , \\
\partial_t p_{010} =& - \Gamma_s ( p_{010} - p_{011} - p_{110} ) , \\
\partial_t p_{001} =& - (2 \Gamma_\uparrow + \Gamma_\downarrow) p_{001} 
+ \Gamma_\uparrow p_{000} + \Gamma_\downarrow ( p_{011} + p_{101} ) , \\
\partial_t p_{100} =& - (2 \Gamma_\uparrow + \Gamma_\downarrow) p_{100} 
+  \Gamma_\uparrow p_{000} + \Gamma_\downarrow ( p_{110} + p_{101} ) , \\
\partial_t p_{000} =& - 2 \Gamma_\uparrow  p_{000} + 
\Gamma_\downarrow ( p_{001} + p_{100} ) + \Gamma_s p_{010} ,
\end{align}
The probability of full excitation $p_{111}$ decays to zero 
on the timescale $t > \Gamma_s^{-1}$. 
The probabilities in the last three equations relax with the very 
fast rates $\sim 2 \Gamma_\uparrow$ and can be adiabatically eliminated. 
Setting there $\partial_t p_{abc} =0$, and substituting the resulting 
solutions in the other equations, we obtain
\begin{align}
\partial_t p_{011} =& - (\Gamma_s + \hlf \Gamma_\downarrow) p_{011} + 
\hlf \Gamma_\downarrow p_{101} + \qtr \Gamma_s p_{010} , \\
\partial_t p_{101} =& - \Gamma_\downarrow p_{101} + \hlf \Gamma_s p_{010} 
+ \hlf \Gamma_\downarrow ( p_{011} + p_{110} ) , \\
\partial_t p_{110} =& - (\Gamma_s + \hlf \Gamma_\downarrow) p_{110} 
+ \hlf \Gamma_\downarrow p_{101} + \qtr \Gamma_s p_{010} , \\
\partial_t p_{010} =& - \Gamma_s p_{010} + \Gamma_s ( p_{011} + p_{110} ) 
\end{align}
where we assumed
$\frac{\Gamma_\downarrow}{\Gamma_s} p_{001}, 
\frac{\Gamma_\downarrow}{\Gamma_s} p_{100} \ll p_{010}$ 
and $\Gamma_\downarrow \ll 2\Gamma_\uparrow$.

By adiabatic elimination of the states with neighboring ground state 
superatoms, we projected the system onto a reduced configuration space.
From the above rate equations, we can deduce three fundamental processes
affecting the holes (see Fig. \ref{fig:EffectiveHoleModel} lower panel): 
\begin{itemize}
\item[(i)] creation of holes with the rate $\Gamma_c = \Gamma_s$, 
\item[(ii)] annihilation of one of the two holes separated by one excitation 
with the rate $\Gamma_a = \Gamma_s$, 
\item[(iii)] transport of holes between neighboring sites with rate
$\Gamma_t = \hlf \Gamma_\downarrow$.
\end{itemize}
The corresponding jump operators, involving the constraints that no two 
neighboring holes are allowed, are given in Sec. \ref{sec:HoleDynamicsSS}.

\bibliography{2017-05-24_SALattFac.bib}

\end{document}